\documentclass[prb,two column]{revtex4}
\usepackage{amssymb}
\usepackage{amsmath}
\usepackage{graphicx}

\begin{document}

\title{The geometrical nature and some properties of the capacitance
coefficients based on the Laplace's equation}
\author{William J. Herrera}
\email{jherreraw@unal.edu.co}
\author{Rodolfo A. Diaz}
\email{radiazs@unal.edu.co}
\affiliation{Departamento de F\'{\i}sica, Universidad Nacional de Colombia, Bogot\'{a}
Colombia}

\begin{abstract}
The fact that the capacitance coefficients for a set of conductors are
geometrical factors is derived in most electricity and magnetism textbooks.
We present an alternative derivation based on Laplace's equation that is
accessible for an intermediate course on electricity and magnetism. The
properties of Laplace's equation permits to prove many properties of the
capacitance matrix. Some examples are given to illustrate the usefulness of
such properties.
\end{abstract}

\maketitle


\section{\label{sec:int}Introduction}

The fact that the capacitance is a geometrical factor is an important
property in courses on electricity and magnetism.\cite{Berkeley, Jack}
Derivations of this property are usually based on the principle of
superposition\cite{Berkeley, Jack} and the Green function formalism\cite%
{Uehara, Lorenzo}. Nevertheless, such derivations are not convenient for
calculations. Alternative techniques to calculate the capacitance
coefficients based on the Green function formalism\cite{Donolato} and other
methods\cite{Cui, Bodegom, Tong} have been developed.

In this paper we give a simple proof of the geometrical nature of the
capacitance coefficients based on Laplace's equation. Our approach permits
to demonstrate many properties of the capacitance matrix. The method is
illustrated by reproducing some well known results, and applications in
complex situations are suggested.

\section{Capacitance coefficients\label{sec:coef}}

\begin{figure}[h]
\begin{center}
\includegraphics[width=6.5cm]{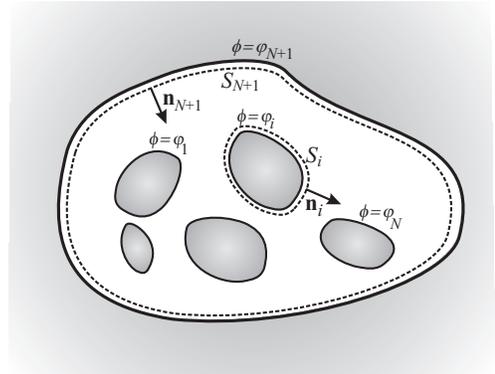}
\end{center}
\caption{A system consisting of $N$ internal conductors with conductor $N+1$
enclosing them. The normals $\mathbf{n}_{i}$ with $i=1,\ldots,N+1$ point
outward with respect to the conductors and inward with respect to the volume 
$V_{S_{T}}$ (defined by the region in white). The surfaces $S_{i}$ with $%
i=1,\ldots,N$ are slightly bigger than the ones corresponding to the
conductors. In contrast, the surface $S_{N+1}$ is slightly smaller than the
surface of the external conductor.}
\label{fig:Ncond}
\end{figure}

We consider a system of $N$ internal conductors and an external conductor
that encloses them. The potential on each internal conductor is denoted by $%
\varphi_{i}$, $i=1,2,\ldots ,N$. The surface of the external conductor is
denoted by $S_{N+1}$, and its potential is denoted by $\varphi_{N+1}$ (see
Fig.~\ref{fig:Ncond}). One reason to introduce the external conductor is
that it provides a closed boundary to ensure the uniqueness of the
solutions. In addition, many capacitors contain an enclosing conductor as
for the case of spherical concentric shells. As we shall see, the case in
which there is no external conductor can be obtained in the appropriate
limit.

The surface charge density $\sigma $ on an electrostatic conductor is given
by\cite{Berkeley, Jack} 
\begin{equation}
\sigma _{i}=\varepsilon _{0}\mathbf{E}\cdot \mathbf{n}_{i}=-\varepsilon
_{0}\nabla \phi \cdot \mathbf{n}_{i}\qquad (i=1,\ldots ,N+1),
\end{equation}%
where $\mathbf{n}_{i}$ is an unit vector normal to the surface $S_{i}$
pointing outward with respect to the conductor (see Fig.~\ref{fig:Ncond}); $%
\mathbf{E}$ and $\phi $ denote the electrostatic field and potential
respectively. The charge on each conductor is given by 
\begin{equation}
Q_{i}=\!\oint_{S_{i}}\sigma _{i}\,dS=-\varepsilon _{0}\!\oint_{S_{i}}\nabla
\phi \cdot \mathbf{n}_{i}\,dS.  \label{Qigen}
\end{equation}%
The surface $S_{i}$ encloses the conductor $i$ and is arbitrarily near and
locally parallel to the real surface of the conductor (see Fig.~\ref%
{fig:Ncond}).\cite{note1} We define the total surface $S_{T}$ as 
\begin{equation}
S_{T}=S_{1}+\ldots +S_{N}+S_{N+1}.  \label{ST}
\end{equation}%
The volume $V_{S_{T}}$ defined by the surface $S_{T}$ is the one delimited
by the external surface $S_{N+1}$ and the $N$ internal surfaces $S_{i}$. The
potential $\phi $ in such a volume must satisfy Laplace's equation with the
boundary conditions 
\begin{equation}
\phi (S_{i})=\varphi _{i}\qquad (i=1,\ldots ,N+1).  \label{cond front fi}
\end{equation}%
Because of the linearity of Laplace's equation, the solution for $\phi $ can
be parameterized as 
\begin{equation}
\phi =\sum_{j=1}^{N+1}\varphi _{j}f_{j},  \label{fi=fi fi}
\end{equation}%
where the $f_{j}$ are functions that satisfy Laplace's equation in the
volume $V_{S_{T}}$ with the boundary conditions 
\begin{equation}
\nabla ^{2}f_{j}=0,\text{ \ \ }f_{j}(S_{i})=\delta _{ij},\;\;(i,j=1,\ldots
,N+1).  \label{cond fi}
\end{equation}%
The solutions for $f_{j}$ ensure that $\phi $ is the solution of Laplace's
equation with the boundary conditions in Eq.~(\ref{cond front fi}). The
uniqueness theorem also ensures that the solution for each $f_{j}$ is unique
(as is the solution for $\phi $). The boundary conditions (\ref{cond fi})
indicate that the $f_{j}$ functions depend only on the geometry.

If we apply the gradient operator in Eq.~(\ref{fi=fi fi}) and substitute the
result into Eq.~(\ref{Qigen}), we obtain 
\begin{subequations}
\label{Q_i_sistema_cargas}
\begin{align}
Q_{i}& =\sum_{j=1}^{N+1}C_{ij}\varphi_{j}  \label{Q_i_sistema_cargas.a} \\
C_{ij}& \equiv -\varepsilon_{0}\oint_{S_{i}}\nabla f_{j}\cdot \mathbf{n}%
_{i}\,dS,  \label{Q_i_sistema_cargas.b}
\end{align}
which shows that the $C_{ij}$ factors are exclusively geometric. The
symmetry of the associated $C_{ij}$ matrix can be obtained by purely
geometrical arguments. We start from the definition of $C_{ij}$ in Eq.~(\ref%
{Q_i_sistema_cargas.b}) and find 
\end{subequations}
\begin{equation}
C_{ij}=-\varepsilon_{0}\!\oint_{S_{i}}\nabla f_{j}\cdot \mathbf{n}%
_{i}\,dS=\varepsilon_{0}\!\oint_{S_{T}}f_{i}\nabla f_{j}\cdot (-\mathbf{n}%
_{i})\,dS,
\end{equation}
where we have used the fact that $f_{i}=1$ on the surface $S_{i}$ and zero
on the other surfaces. From Gauss's theorem we obtain 
\begin{subequations}
\begin{align}
C_{ij}& =\varepsilon_{0}\int_{V_{S_{T}}}\nabla \cdot (f_{i}\nabla f_{j})\,dV
\\
& =\varepsilon_{0}\int_{V_{S_{T}}}[\nabla f_{i}\cdot \nabla
f_{j}+f_{i}\nabla^{2}f_{j}]\,dV.
\end{align}
Because $\nabla^{2}f_{j}=0$ in $V_{S_{T}}$, it follows that 
\end{subequations}
\begin{equation}
C_{ij}=\varepsilon_{0}\!\int_{V_{S_{T}}}\nabla f_{i}\cdot \nabla f_{j}\,dV,
\label{Cij 2}
\end{equation}
Equation~\eqref{Cij 2} implies that $C_{ij}$ is symmetric,\cite{note2} that
is, 
\begin{equation}
C_{ij}=C_{ji}.  \label{Prop2}
\end{equation}

For certain configuration of conductors, consider two sets of charges and
potentials $\{Q_{i},\varphi _{i}\}$ and $\{Q_{i}^{\prime },\varphi
_{i}^{\prime }\}$. From Eqs.~(\ref{Q_i_sistema_cargas}) and (\ref{Prop2}) we
have that 
\begin{subequations}
\begin{align}
\sum_{i=1}^{N+1}Q_{i}\varphi _{i}^{\prime }&
=\sum_{i=1}^{N+1}(\sum_{j=1}^{N+1}C_{ij}\varphi _{j})\varphi _{i}^{\prime }
\\
& =\sum_{j=1}^{N+1}(\sum_{i=1}^{N+1}C_{ji}\varphi _{i}^{\prime })\varphi
_{j},
\end{align}%
which implies that 
\end{subequations}
\begin{equation}
\sum_{i=1}^{N+1}Q_{i}\varphi _{i}^{\prime }=\sum_{j=1}^{N+1}Q_{j}^{\prime
}\varphi _{j}.  \label{recip}
\end{equation}%
Equation (\ref{recip}) is known as the reciprocity theorem.\cite{Berkeley}

When one or more of the $N$ internal conductors has an empty cavity, is well
known that there is no charge induced on the surface of the cavity \cite%
{Berkeley,Jack} (let us call it $S_{i,\,c}$). Consequently, although $%
S_{i,\,c} $ is part of the surface of the conductor, such a surface can be
excluded in the integration in Eq.~(\ref{Qigen}). In addition, we can check
by uniqueness that $f_{j}=\delta_{ij}$ in the volume of the cavity $%
V_{i,\,c} $ so that $\nabla f_{j}=0$ in such a volume, and hence it can be
excluded from the volume integral (\ref{Cij 2}). In conclusion neither $%
S_{i,\,c}$ nor $V_{i,\,c}$ contribute in this case.

\begin{figure}[h]
\begin{center}
\includegraphics[width=6.5cm]{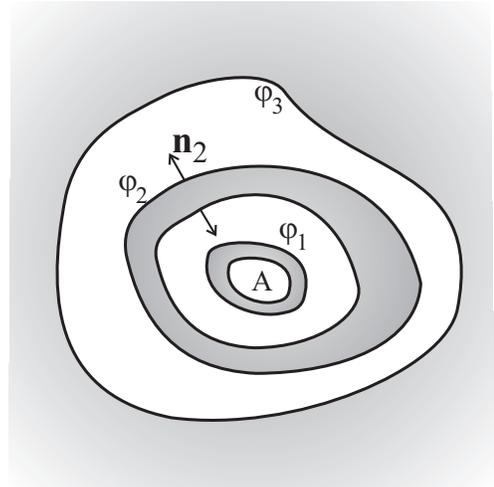}
\end{center}
\caption{Example of system in which there is a successive embedding of
conductors. The volume $V_{S_{T}}$ corresponds to the region in white. The
regions corresponding to empty cavities (and their associated surfaces and
volumes) can be excluded without affecting the calculations. In this picture
cavity A is empty and its surface and volume need not be considered for
calculations.}
\label{fig:Ncond2}
\end{figure}

The situation is different if there is another conductor in the cavity. In
this case, the surface of the cavity contributes in Eq.~(\ref{Qigen}).
Similarly the volume between the cavity and the embedded conductor
contributes in the volume integral (\ref{Cij 2}). The arguments can be
extended for successive embedding of conductors in cavities as shown by Fig.~%
\ref{fig:Ncond2} or for conductors with several cavities.

\section{Some additional properties}

We define a function $F$ 
\begin{equation}
F\equiv \sum_{j=1}^{N+1}f_{j},
\end{equation}%
and see from Eq.~(\ref{cond fi}) that 
\begin{equation}
\nabla ^{2}F=0,\qquad F(S_{i})=1\qquad (i=1,\ldots ,N+1).
\end{equation}%
Since $F=1$ throughout the surface $S_{T}$, we see by uniqueness that $F=1$
in the volume $V_{S_{T}}$ from which we find that 
\begin{equation}
\sum_{j=1}^{N+1}f_{j}=1.  \label{Propf}
\end{equation}%
In addition, by summing over $j$ in Eq.~(\ref{Q_i_sistema_cargas.b}) and
taking into account Eq.~(\ref{Propf}), we find that 
\begin{equation}
\sum_{j=1}^{N+1}C_{ij}=0\qquad (i=1,\ldots ,N+1).  \label{Cij j}
\end{equation}%
The symmetry of the $C_{ij}$ elements leads also to 
\begin{equation}
\sum_{i=1}^{N+1}C_{ij}=0\qquad (j=1,\ldots ,N+1).  \label{Prop1}
\end{equation}

Equations~\eqref{Cij j} and \eqref{Prop1} imply that the sum of the elements
over any row or column of the matrix is zero. Appendix~\ref{ap:pruebas}
gives some proofs of consistency for these important properties. Taking into
account the symmetrical nature of the $C_{ij}$ matrix with dimensions $%
(N+1)\times (N+1)$ and the $N+1$ constraints in Eq.~(\ref{Prop1}), we see
that for a system of $N$ conductors surrounded by another conductor $N+1$,
the number of independent capacitance coefficients is 
\begin{equation}
(N+1)^{2}-\left[ \frac{N(N+1)}{2}\right] -(N+1)=\frac{N(N+1)}{2},
\label{grad lib}
\end{equation}

Other important properties are that 
\begin{subequations}
\label{Prop3}
\begin{align}
C_{ii}& \geq 0  \label{Prop3.a} \\
C_{ij}& \leq 0,\qquad (i\neq j).  \label{Prop3.b}
\end{align}
Equation~\eqref{Prop3.a} follows straightforwardly from Eq.~(\ref{Cij 2}).
To demonstrate Eq.~\eqref{Prop3.b}, we recall that the solutions of
Laplace's equation cannot have local minima nor local maxima in the volume
in which the equation is valid.\cite{Berkeley, Jack} Consequently, the $%
f_{j} $ functions must lie in the interval 
\end{subequations}
\begin{equation}
0\leq f_{j}\leq 1.  \label{Prop fj}
\end{equation}
Because $f_{j}=0$ on any surface $S_{i}$ for $i\neq j$, we see that $f_{j}$
acquires its minimum value on such surfaces. Therefore the function $\nabla
f_{j}$ should point outward with respect to the conductor $i$ for $i\neq j$.
Hence 
\begin{equation}
\mathbf{n}_{i} \cdot \nabla f_{j} \geq 0\ \mbox{for}\ i\neq j.  \label{gradf}
\end{equation}
We substitute Eq.~(\ref{gradf}) into Eq.~(\ref{Q_i_sistema_cargas}) and
obtain $C_{ij}\leq 0$ for $i\neq j$. An additional derivation of the fact
that $C_{ii}\geq 0$ can be obtained by taking into account that $f_{j}$
acquires its maximum value on the surface $S_{j}$.

Equation~(\ref{Prop1}) can be rewritten as 
\begin{equation}
\sum_{i=1}^{N}C_{ij}=-C_{N+1,j}.
\end{equation}
From Eq.~(\ref{Prop3}) we have that $C_{N+1,j}\leq 0$ for $j=1,\ldots ,N$
and $C_{N+1,N+1}\geq 0$. Hence 
\begin{subequations}
\label{Prop4all}
\begin{align}
\sum_{i=1}^{N}C_{ij}& \geq 0\qquad (j=1,\ldots ,N)  \label{Prop4} \\
\sum_{i=1}^{N}C_{i,N+1}& \leq 0.  \label{Prop4a}
\end{align}
The following properties follow from Eqs.~(\ref{Prop2}), \eqref{Prop1}, %
\eqref{Prop3}, and \eqref{Prop4all} 
\end{subequations}
\begin{subequations}
\label{prop5all}
\begin{align}
|C_{ii}|& \geq \sum_{i\neq j}^{N}|C_{ij}| \\
|C_{ii}|& \geq |C_{ij}|, \\
C_{ii}C_{jj}& \geq C_{ij}^{2}  \label{Prop5} \\
|C_{N+1,N+1}|& =\sum_{i=1}^{N}|C_{i,N+1}| \\
|C_{N+1,N+1}|& \geq |C_{i,N+1}|,  \label{Prop5a}
\end{align}
where $i,j=1,\ldots ,N$.

A particularly interesting case arises when the external conductor is at
zero potential. In such a case, although the elements of the form $C_{N+1,j}$
do not necessarily vanish, they do not appear in the contributions to the
charge on the internal conductors as can be seen from Eq.~(\ref%
{Q_i_sistema_cargas}) by setting $\varphi_{N+1}=0$. For this reason, the
capacitance matrix used to describe $N$ free conductors (that is, not
surrounded by another conductor) has dimensions $N\times N$.\cite{note3}

\section{Two conductors}

We illustrate our method by deriving the basic properties of a system of two
conductors. These examples will show the usefulness of Eq.~(\ref%
{Q_i_sistema_cargas}) and some of the properties derived from our approach.
We analyze a single internal conductor with an external conductor that is, $%
N=1$. The internal conductor is labeled as conductor 1. From Eqs.~(\ref%
{Prop2}) and (\ref{Prop1}) we have 
\end{subequations}
\begin{equation}
C_{21}=C_{12}=-C_{11}=-C_{22}.
\end{equation}
Therefore, there is only one independent coefficient, say $C_{11}$ (in
agreement with Eq.~\eqref{grad lib} with $N=1$). The charges on the internal
and external conductors can be calculated from Eq.~(\ref{Q_i_sistema_cargas}%
) 
\begin{subequations}
\begin{align}
Q_{1}& =C_{11}(\varphi_{1}-\varphi_{2}) \\
Q_{2}& =-C_{11}(\varphi_{1}-\varphi_{2})=-Q_{1}.  \label{27b}
\end{align}
Equation~\eqref{27b} is consistent with Eq.~(\ref{ext int}) and shows that
the charge induced on the surface of the cavity of the conductor 2 is
opposite to the charge on the conductor 1.

In Table~\ref{tab:resumen} we display the results of three well known
configurations of two conductors. The second column shows the $f_{i}$
functions, which can be found by Laplace's equation (\ref{cond fi}) and used
to calculate $C_{11}$ with Eq.~(\ref{Q_i_sistema_cargas}).

\begin{table}[h]
\begin{center}
\begin{tabular}{|p{5cm}|c|c|}
\hline
System & $f_{1}$ & $C_{11}$ \\ \hline
Spherical shell with radius $b$ and concentric solid sphere with radius $a$.
& $\frac{ab}{b-a}\left( \frac{1}{r}-\frac{1}{b}\right) $ & $\frac{4\pi
\varepsilon _{0}ab}{b-a}$ \\ \hline
Cylindrical shell with radius $b$ and concentric solid cylinder with radius $%
a$, both with length $L$. & $\frac{\ln (r/b)}{\ln (a/b)}$ & $\frac{2\pi
\varepsilon _{0}L}{\ln (b/a)}$ \\ \hline
Two parallel planes with area $A$ at $x=0$ and $x=d$ (conductor 1). & $x/d$
& $\varepsilon _{0}\frac{A}{d}$ \\ \hline
\end{tabular}%
\end{center}
\caption{$C_{11}$ and $f_{1}$ factors for three systems of two conductors
with $a$ $\leq r\leq b$ and $0\leq x\leq d$. We neglect edge effects for the
cylinders and planes.}
\label{tab:resumen}
\end{table}

\section{Examples}

We use our approach to study a system with embedded of conductors. In
addition, the case of two internal conductors is examined, and we show the
limit in which the configuration of two conductors without external
conductor is obtained. These examples show how the properties we have
derived can be used to calculate the capacitance coefficients.

\textit{Example 1}. Consider two concentric spherical shells with radii $b$
and $c$ and a solid spherical conductor (concentric with the others) with
radius $a$ such that $c>b>a$. The potentials are denoted by $\varphi _{1}$, $%
\varphi_{2}$, and $\varphi_{3}$ respectively. The general solution of
Laplace's equation for $f_{i}$ can be written as 
\end{subequations}
\begin{equation}
f_{i}=\frac{A_{i}}{r}+B_{i}.  \label{f param}
\end{equation}
From Eqs. (\ref{cond fi}) and (\ref{f param}) we obtain $f_{1}$ and $f_{3}$ 
\begin{equation}
f_{1}= 
\begin{cases}
\frac{ab}{b-a}\Big(\frac{1}{r}-\frac{1}{b}\Big) & (a\leq r\leq b) \\ 
0 & (b\leq r\leq c)%
\end{cases}%
\end{equation}
\begin{equation}
f_{3}= 
\begin{cases}
0 & (a\leq r\leq b) \\ 
\frac{bc}{c-b}\left( \frac{1}{b}-\frac{1}{r}\right) & (b\leq r\leq c)%
\end{cases}
.
\end{equation}
Although $f_{2}$ can be obtained the same way, it is easier to extract it
from Eq.~(\ref{Propf}). The result is 
\begin{equation}
f_{2}= 
\begin{cases}
\frac{ab}{b-a}\left( \frac{1}{a}-\frac{1}{r}\right) & (a\leq r\leq b) \\ 
\frac{bc}{c-b}\left( \frac{1}{r}-\frac{1}{c}\right) & (b\leq r\leq c)%
\end{cases}
.  \label{f2}
\end{equation}
The nine capacitance coefficients can be evaluated explicitly from Eq.~(\ref%
{Q_i_sistema_cargas}), but it is easier to use Eqs.~(\ref{Prop2}) and (\ref%
{Prop1}) and to take into account that $C_{31}=0$ ($\nabla f_{1}(r)=0$ for $%
r>b$). We have 
\begin{subequations}
\label{Prop_3cap_contenidos}
\begin{align}
C_{13}& =0, & &C_{12}=-C_{11} \\
C_{22}& =C_{11}-C_{32}, & &C_{33}=-C_{32}.
\end{align}

From Eq.~(\ref{Q_i_sistema_cargas}) the charge on each conductor is 
\end{subequations}
\begin{subequations}
\label{Sol_3cond_contenidos}
\begin{align}
Q_{1}& =C_{11}(\varphi _{1}-\varphi _{2}) \\
Q_{2}& =-Q_{1}+C_{32}(\varphi _{3}-\varphi _{2}) \\
Q_{3}& =C_{32}(\varphi _{2}-\varphi _{3})=-(Q_{1}+Q_{2}).
\end{align}%
Hence, we only have to calculate $C_{11}$ and $C_{32}$. \cite{note4} The
result gives 
\end{subequations}
\begin{equation}
C_{11}=4\pi \varepsilon _{0}\frac{ab}{b-a},\quad C_{32}=-4\pi \varepsilon
_{0}\frac{bc}{c-b}.
\end{equation}

If $\varphi _{2}=\varphi _{3}$, we find that $Q_{1}=-Q_{2}$ and $Q_{3}=0$.
It can be shown that Eqs.~(\ref{Prop_3cap_contenidos}) and (\ref%
{Sol_3cond_contenidos}) are valid even if the conductors are neither
spherical nor concentric, because those equations come from Eqs.~%
\eqref{Q_i_sistema_cargas.a}, \eqref{Prop2}, and \eqref{Prop1} which are
general properties independent of specific geometries.

\textit{Example 2}. Consider two internal conductors and a grounded external
conductor. As customary, we begin with $Q_{1}=Q_{2}=0$. By transfering
charge from one internal conductor to the other we keep $Q_{1}=-Q_{2}$. From
Eq. (\ref{Q_i_sistema_cargas.a}) and defining $V\equiv \varphi _{1}-\varphi
_{2}\ $we find%
\begin{eqnarray}
Q_{1} &=&\left( C_{11}+C_{12}\right) \varphi _{1}-C_{12}V, \\
Q_{1} &=&-C_{13}\varphi _{1}-C_{12}V,  \label{Q1C}
\end{eqnarray}%
where we have used Eq. (\ref{Prop1}). Similarly $Q_{2}=-C_{23}\varphi
_{1}-C_{22}V$, and using again Eq. (\ref{Prop1}) we find%
\begin{equation}
Q_{1}+Q_{2}=C_{33}\varphi _{1}-C_{32}V.
\end{equation}%
Since the system is neutral $Q_{1}+Q_{2}=0$ and hence%
\begin{equation}
\varphi _{1}=-\frac{C_{32}}{C_{33}}V,  \label{fi1V}
\end{equation}%
substituting Eq. (\ref{fi1V}) into Eq. (\ref{Q1C}) we obtain%
\begin{equation}
Q_{1}=CV\ \ \ ;\ \ \ C\equiv \frac{C_{13}C_{32}-C_{33}C_{12}}{C_{33}}.
\label{Q1=CV}
\end{equation}

Because $N=2$ only three of the coefficients in the definition of $C$ are
independent. From Eqs. \eqref{Prop3} we see that this effective capacitance
is non negative. The procedure is not valid if $C_{33}=0$, in that case we
see by using Eqs. (\ref{Prop1}) and \eqref{Prop3} that $C_{i3}=C_{3i}=0$,
and from Eq. (\ref{Q1C}) we find $C=-C_{12}=C_{22},$ which is also non
negative. The limit in which there is no external conductor is obtained by
taking all the dimensions of the cavity to infinity while keeping the
external conductor grounded as discussed in Ref.~\onlinecite{note3}.

\section{Conclusions}

We have used an approach based on Laplace's equation to demonstrate that the
capacitance matrix depends only on purely geometrical factors. The explicit
use of Laplace's equation permits us to demonstrate many properties of the
capacitance coefficients. The geometrical relations and properties shown
here permits us to simplify many calculations of the capacitance
coefficients. We emphasize that Laplace's equations necessary for finding
the capacitance coefficients are purely geometrical as can be seen from Eqs.~%
\eqref{cond fi} and \eqref{Q_i_sistema_cargas}. Laplace's equation is
usually easier than Green function formalism for either analytical or
numerical calculations. Appendix \ref{ap:pruebas} shows some proofs of
consistency to enhance the physical insight and the reliability of our
method.

\appendix

\section{Proofs of consistency\label{ap:pruebas}}

A proof of consistency for the identity (\ref{Prop1}), is achieved by using
Eq.~(\ref{Q_i_sistema_cargas}) to calculate the total charge on the $N$
internal conductors\cite{note5} 
\begin{equation}
Q_{\mathrm{int}}=\sum_{i=1}^{N}Q_{i}=\sum_{j=1}^{N+1}\left[ \varphi
_{j}\sum_{i=1}^{N}C_{ij}\right] .  \label{Qint}
\end{equation}%
We use Eq.~(\ref{Prop1}) to find 
\begin{equation}
Q_{\mathrm{int}}=-\sum_{j=1}^{N+1}C_{N+1,j}\varphi _{j}.  \label{Qint int}
\end{equation}%
Note that Eq. (\ref{Qint int}) requires many fewer elements of the $C_{ij}$
matrix than Eq.~(\ref{Qint}). This difference becomes more significant as $N$
increases. If we again use Eq.~(\ref{Q_i_sistema_cargas}), we can find the
charge on the cavity of the external conductor 
\begin{equation}
Q_{N+1}=\sum_{j=1}^{N+1}C_{N+1,j}\varphi _{j},
\end{equation}%
and therefore 
\begin{equation}
Q_{N+1}=-Q_{\mathrm{int}},  \label{ext int}
\end{equation}%
a property that can also be obtained from Gauss's law. \cite{Berkeley, Jack}

Another proof of consistency for Eq.~(\ref{Prop1}) is found by employing
Eqs.~(\ref{Q_i_sistema_cargas}) and (\ref{Qint int}) to calculate $Q_{%
\mathrm{int}}$ (taking into account that Eq.~\eqref{Qint int} comes directly
from Eq.~\eqref{Prop1}) 
\begin{subequations}
\begin{align}
Q_{\mathrm{int}}& =-\sum_{j=1}^{N+1}C_{N+1,j}\varphi_{j} \\
& =\varepsilon_{0}\oint_{S_{N+1}}\nabla \left( \sum_{j=1}^{N+1}f_{j}\varphi
_{j}\right) \cdot \mathbf{n}_{N+1}\,dS.
\end{align}
We utilize Eq.~(\ref{fi=fi fi}) to write $Q_{\mathrm{int}}$ as 
\end{subequations}
\begin{equation}
Q_{\mathrm{int}}=\varepsilon_{0}\!\oint_{S_{N+1}}\nabla \phi \cdot \mathbf{n}%
_{N+1}\,dS=\varepsilon_{0}\oint_{S_{N+1}}\mathbf{E}\cdot (-\mathbf{n}%
_{N+1})\,dS.
\end{equation}
This relation is clearly correct because $\mathbf{n}_{N+1}$ points inward
with respect to the volume $V_{S_{T}}$.

A proof of consistency for Eq.~(\ref{Cij 2}) that shows the symmetry of $%
C_{ij}$ can be obtained by calculating the electrostatic internal energy,
which in terms of the electric field is 
\begin{subequations}
\begin{align}
U& =\frac{\varepsilon_{0}}{2}\!\int_{V_{S_{T}}}E^{2}\,dV=\frac{\varepsilon
_{0}}{2}\int_{V_{S_{T}}}\mathbf{\nabla }\phi \cdot \mathbf{\nabla }\phi \,dV
\\
& =\frac{1}{2}\sum_{i,j}^{N+1}\varphi_{i}\varphi_{j}\left[ \varepsilon
_{0}\int_{V_{S_{T}}}\mathbf{\nabla }f_{i}\cdot \mathbf{\nabla }f_{j}\,dV%
\right] ,
\end{align}
where we have used Eq.~(\ref{fi=fi fi}). From Eq.~(\ref{Cij 2}) we find 
\end{subequations}
\begin{equation}
U=\frac{1}{2}\sum_{i,j}^{N+1}C_{ij}\varphi_{j}\varphi_{i}=\frac{1}{2}%
\sum_{i}^{N+1}Q_{i}\varphi_{i},
\end{equation}
consistent with standard results.\cite{Berkeley,Jack}

\section{Suggested Problems}

To enhance the understanding of this approach and its advantages, we give
some general suggestions for the reader.

\begin{enumerate}
\item Implement a numerical method to solve the Laplace's equation (\ref%
{cond fi}) for the $f_{i}$ functions associated with a nontrivial geometry
(for example, two non-concentric ellipsoids). Use Eqs.~(\ref{Propf}) and %
\eqref{Prop fj} to either simplify your calculations or to check the
consistency of your results. Then use Eq.~(\ref{Q_i_sistema_cargas}) to
obtain the $C_{ij}$ factors numerically. Use Eq. (\ref{Prop2}) and Eqs.~%
\eqref{Cij j}--\eqref{prop5all} either to simplify your calculations or to
check the consistency of your results.

\item We have emphasized that to calculate the total charge on the internal
conductors Eq.~(\ref{Qint int}) requires many fewer $C_{ij}$ elements than
Eq.~(\ref{Qint}). How many fewer elements are required for an arbitrary
value of $N$?

\item For a successive embedding of concentric spherical shells, calculate
the capacitance coefficients for an arbitrary number of spheres.

\item Show that for the successive embedding of three conductors with
arbitrary shapes, Eqs.~\eqref{Prop_3cap_contenidos} and %
\eqref{Sol_3cond_contenidos} still hold. Generalize your results for an
arbitrary number of conductors.
\end{enumerate}

\begin{acknowledgments}
We acknowledge the useful suggestions of two anonymous referees. We also
thank Divisi\'{o}n de Investigaci\'{o}n de Bogot\'{a} (DIB) for its
financial support.
\end{acknowledgments}


\begin{thebibliography}{99}
\bibitem{Berkeley} W. Taussig Scott, \emph{The Physics of Electricity and
Magnetism} (John Wiley \& Sons, New York, 1966), 2nd ed.; Gaylord P.
Harnwell, \emph{Principles of Electricity and Electromagnetism}
(McGraw-Hill, New York, 1949); Leigh Page and Norman I. Adams Jr., \emph{%
Principles of Electricity} (D. Van Nostrand, New Jersey, 1958), 3rd ed.; A.
N. Matveev, \emph{Electricity and Magnetism} (Mir, Moscow, 1988).

\bibitem{Jack} J. D. Jackson, \emph{Classical Electrodynamics} (John Wiley
\& Sons, New Jersey, 1998), 3rd ed.; David J. Griffiths, \emph{Introduction
to Electrodynamics} (Prentice Hall, New Jersey, 1999), 3rd ed.

\bibitem{Uehara} Mituo Uehara, ``Green's functions and coefficients of
capacitance,'' Am. J. Phys. \textbf{54} (2), 184--185 (1986).

\bibitem{Lorenzo} V. Lorenzo and B. Carrascal, ``Green's functions and
symmetry of the coefficients of a capacitance matrix,'' Am. J. Phys. \textbf{%
56} (6), 565 (1988).

\bibitem{Donolato} C. Donolato, ``Approximate evaluation of capacitances by
means of Green's reciprocal theorem,'' Am. J. Phys. \textbf{64} (8),
1049--1054 (1996).

\bibitem{Cui} Y. Cui, ``A simple and convenient calculation of the
capacitance for an isolated conductor plate,'' Eur. J. Phys. \textbf{17}
(6), 363--364 (1996).

\bibitem{Bodegom} E. Bodegom and P. T. Leung, ``A surprising twist to a
simple capacitor problem,'' Eur. J. Phys. \textbf{14} (2), 57--58 (1993).

\bibitem{Tong} G. P. Tong, ``Electrostatics of two conducting spheres
intersecting at angles,'' Eur. J. Phys. \textbf{17} (4), 244--249 (1996).

\bibitem{note1} $Q_{N+1}$ is not necessarily the total charge on the
external conductor, but the charge accumulated on the surface of the cavity
that encloses the other conductors. The value of the charge is calculated
with the surface integral (\ref{Qigen}), which for the case of the internal
conductors encompasses the whole surface, but for the external conductor is
only the surface of the cavity that encloses the other conductors.

\bibitem{note2} Equation~(\ref{Cij 2}) is an integral over the volume for
the $C_{ij}$ factors. We might be tempted to use Gauss' theorem to obtain an
integral of the volume directly from Eq.~(\ref{Q_i_sistema_cargas}).
However, $f_{j}$ is not defined in the region inside the conductors. The
gradient of $f_{j}$ in Eq.~(\ref{Q_i_sistema_cargas}) is evaluated in an
external neighborhood of the conductor surface.

\bibitem{note3} By uniqueness, the solution for this problem is equivalent
to the solution for a system consisting of the same $N$ conductors contained
in the cavity of a surrounding conductor, such that all the dimensions of
the cavity tend to infinity, and the potential of the external conductor is
set to zero.

\bibitem{note4} If we take into account that $C_{13}$ is another degree of
freedom (although zero), we have a total of three degrees of freedom, in
agreement with Eq.~\eqref{grad lib} for $N=2$.

\bibitem{note5} For a derivation of some of these results based on the
energy of the electrostatic field see L. D. Landau, E. M. Lifshitz, and L.
P. Pitaevskii, \emph{Electrodynamics of Continuous Media} (Elsevier
Butterworth-Heinemann, 1984), 2nd ed., p. 3.
\end{thebibliography}
\end{document}